\documentclass[prd,showpacs,twocolumn,preprintnumbers,amsmath,amssymb,superscriptaddress,floatfix,nofootinbib]{revtex4-2}%

\usepackage{graphicx}
\usepackage{bm}
\usepackage{amsmath}
\usepackage{amsfonts}
\usepackage{amssymb}
\usepackage{color}
\usepackage{multirow}
\usepackage{subfigure}
\usepackage{ulem} 
\usepackage[colorlinks, citecolor=blue,anchorcolor=red,menucolor=red, linkcolor=red,filecolor=red,runcolor=red,urlcolor=blue,frenchlinks=red]{hyperref}

\begin{document}
 
    \title{Unveiling the elusive $\Sigma(1380)$ resonance through coupled-channel dynamics in $\Lambda_c^+\to\eta\pi^+\Lambda$ reaction}
    
    
    
    
    
    

	\author{Wen-Tao Lyu}
	\affiliation{School of Physics, Zhengzhou University, Zhengzhou 450001, China}
	\affiliation{Departamento de Física Teórica and IFIC, Centro Mixto Universidad de Valencia-CSIC Institutos de Investigación de Paterna, 46071 Valencia, Spain}
	
	\author{Si-Wei Liu}
	\affiliation{State Key Laboratory of Heavy Ion Science and Technology, Institute of Modern Physics, Chinese Academy of Sciences, Lanzhou 730000, China} 
    \affiliation{School of Nuclear Sciences and Technology, University of Chinese Academy of Sciences, Beijing 101408, China}

	\author{Jia-Jun Wu}\email{wujiajun@ucas.ac.cn}
	\affiliation{School of Physical Sciences, University of Chinese Academy of Sciences, Beijing 100049, China}
	\affiliation{Southern Center for Nuclear-Science Theory (SCNT), Institute of Modern Physics, Chinese Academy of Sciences, Huizhou 516000, Guangdong Province, China}\vspace{0.5cm}
 
	\author{De-Min Li}\email{lidm@zzu.edu.cn}
	\affiliation{School of Physics, Zhengzhou University, Zhengzhou 450001, China}\vspace{0.5cm}
 
	\author{En Wang}\email{wangen@zzu.edu.cn}
	\affiliation{School of Physics, Zhengzhou University, Zhengzhou 450001, China}\vspace{0.5cm}

\begin{abstract}
We investigate the $\Lambda_c^+ \to \eta \pi^+ \Lambda$ decay measured by the Belle and BESIII Collaborations, focusing on the possible role of the $\Sigma(1380)$ state with spin-parity $J^P=1/2^-$.
In our theoretical framework, the $\Lambda(1670)$ and $a_0(980)$ are dynamically generated from meson-baryon and meson-meson final-state interactions, respectively, 
and the corresponding line shapes of these two states used here are applicable to all relevant hadronic reactions.
Furthermore, the contributions from the intermediate $\Sigma(1385)$ resonance and the possible $\Sigma(1380)$ state are included explicitly. 
By comparing the invariant mass and angular distributions obtained with and without the $\Sigma(1380)$ state, we demonstrate that this state plays an important role in improving the description of the experimental data. 
We also identify the kinematic regions most sensitive to the possible $\Sigma(1380)$ contribution. 
Future high-precision measurements of this process will be instrumental for testing the existence of the $\Sigma$ state with $J^P=1/2^-$.
\end{abstract}
	
\date{\today}
	
\maketitle
	
\section{Introduction}\label{sec1}
The ground-state octet and decuplet baryons are well established, but the nature of the low-lying excited $\Sigma$ baryon with  $I(J^P)=1(1/2^-)$ remains unsettled. 
In the Review of Particle Physics (RPP)~\cite{ParticleDataGroup:2024cfk}, the $J^P=1/2^-$ state $\Sigma(1620)$ is listed with only a one-star rating, indicating that its existence requires further experimental confirmation. 
In addition, chiral unitary approaches predict a dynamically generated $\Sigma(1430)$ state with spin-parity $J^P=1/2^-$ from $S$-wave pseudoscalar meson-octet baryon scattering in the strangeness $S = -1$ sector, with a mass close to the $\bar{K}N$ threshold~\cite{Oset:2001cn,Oset:1997it,Khemchandani:2018amu,Kamiya:2016jqc,Jido:2003cb,Oller:2006jw,Garcia-Recio:2002yxy,Lutz:2001yb}. 
Furthermore, a reexamination of the $\gamma p \to K\Sigma\pi$ reaction using CLAS data indicates a potential signal near 1430~MeV~\cite{CLAS:2013rjt,Roca:2013cca}. 
To further constrain the properties of the $\Sigma(1/2^-)$ state, a wide array of alternative processes has been proposed~\cite{Dai:2018hqb,Xie:2018gbi,Wang:2015qta,Liu:2017hdx,Li:2025exm,Zhang:2026igc,Lyu:2023oqn,Kim:2021wov,Zou:2026ngp,Yao:2025qor,Ma:2025wbz,Lin:2025pyk,Li:2024tvo}. 
Belle has observed a signal close to the $\bar{K}N$ threshold in the $\Lambda\pi$ invariant mass distribution of the process $\Lambda_c^+\to \Lambda \pi^+\pi^+\pi^-$~\cite{Belle:2022ywa}, which could be associated with the $\Sigma(1/2^-)$~\cite{Li:2025yad,He:2025vij}.
For a detailed and up-to-date discussion on the status of this elusive state, we refer the reader to the recent review in Ref.~\cite{Wang:2024jyk}.

On the other hand, several theoretical and phenomenological studies suggest the existence of a lower mass $I(J^P)=1(1/2^-)$ candidate around $1380$~MeV (denoted as $\Sigma(1380)$)~\cite{Zhang:2004xt,Lu:2022hwm}. 
For example, the $\Sigma(1380)$ has been suggested to play an important role in the reactions $K^-p\to \Lambda\pi^+\pi^-$~\cite{Wu:2009nw,Wu:2009tu}, $\Lambda p \to \Lambda p \pi^0$~\cite{Xie:2014zga}, $\gamma N \rightarrow K^{+}\Sigma^{*}(1385) \rightarrow K^{+}\pi\Lambda$~\cite{Chen:2013vxa}, $\Lambda_c^+\to\gamma\pi^+\Lambda$~\cite{Wang:2024ewe}, $\Lambda_c^+ \to p \bar{K}^0\eta$~\cite{Li:2024rqb}, and $K\Sigma^*$ photoproduction~\cite{Gao:2010hy}.
To further investigate this resonance, the $\Lambda_c^+\to\eta\Lambda\pi^+$ decay was proposed as a promising channel to search for the $\Sigma(1380)$ state~\cite{Xie:2017xwx,Wang:2022nac}. 
The Belle Collaboration analyzed this process in 2021~\cite{Belle:2020xku}. 
Subsequently, Ref.~\cite{Lyu:2024qgc} studied this decay and emphasized its potential for probing the low-lying state $\Sigma(1380)$. 
More recently, the BESIII Collaboration analyzed this process and reported evidence for the $\Sigma(1380)$ with $J^P=1/2^-$ at a statistical significance exceeding $3\sigma$~\cite{BESIII:2024mbf,Wang:2026daa}. 
However, a subsequent study argued that the BESIII data could be described without introducing the $\Sigma(1380)$ contribution~\cite{Duan:2024czu}. 
Nevertheless, visible deviations between the calculation and the data remain in the low energy region of the $\pi^+ \Lambda$ invariant mass distribution.
This unresolved tension motivates a reanalysis of this decay with careful consideration of the possible $\Sigma(1380)$ contribution. 

In the present work, we investigate the role of the $\Sigma(1380)$ state in the $\Lambda_c^+\to\eta\Lambda\pi^+$ decay and identify key observables that are sensitive to its contribution. 
In addition to the $\Sigma(1380)$, our model incorporates the $\Lambda(1670)$ generated dynamically from meson-baryon interactions, the $a_0(980)$ generated from meson-meson interactions, and the intermediate $\Sigma(1385)$ resonance. 
We analyze the Monte Carlo (MC) sample provided by the BESIII Collaboration together with the Belle data~\cite{Belle:thesis}, and present detailed results for the $\eta\Lambda$, $\pi^+\eta$, and $\pi^+\Lambda$ invariant mass distributions, as well as the corresponding angular distributions.

Another motivation for our detailed re-examination of this process is that the formulas employed in experimental data analysis are in urgent need of modernization~\cite{Lyu_private}. 
In the data analyses conducted by BESIII~\cite{BESIII:2024mbf} and Belle~\cite{Belle:2020xku}, the conventional Flatté and Breit-Wigner (BW) parametrizations have been used for the $a_0(980)$ and $\Lambda(1670)$ resonances, respectively. 
When applied to high-statistics experimental data, the inherent limitations of these parametrizations, such as their failure to satisfy unitarity and the large number of free parameters, can compromise the quality of the data description. 
This issue is further exacerbated by the fact that different line shapes have been adopted in many analyses of the $a_0(980)$ resonance~\cite{Lyu_private}.
Therefore, in this work we aim to build upon our previous descriptions of hadronic resonances with strong coupled-channel interactions, including the $a_0(980)$ and $\Lambda(1670)$. 
These descriptions feature fewer free parameters and more universal line shapes, which will reduce systematic uncertainties when extracting resonance information from the data.

This paper is organized as follows. In Sec.~\ref{sec2}, we present the theoretical formalism for the $\Lambda_c^+\to\eta\pi^+\Lambda$ decay. The numerical results and corresponding discussions are provided in Sec.~\ref{sec3}. Finally, a brief summary is given in Sec.~\ref{sec4}.

\section{Formalism}\label{sec2}

\begin{figure*}[htbp]
 \subfigure[]{
		\centering
		\includegraphics[scale=0.33]{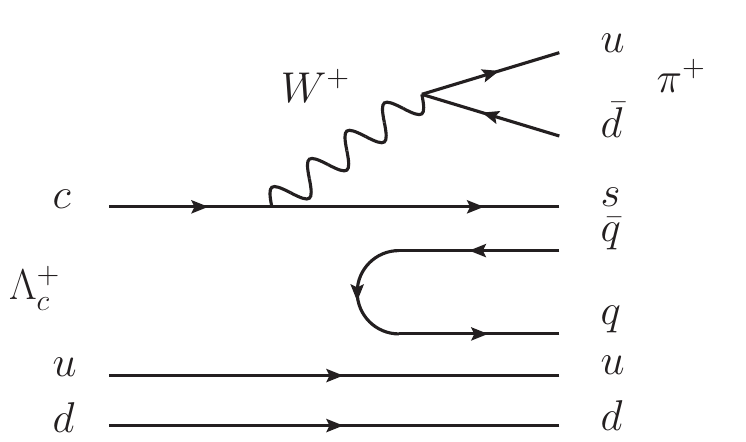}\label{fig:quark_level_a}
	}
	\subfigure[]{
		\centering
		\includegraphics[scale=0.33]{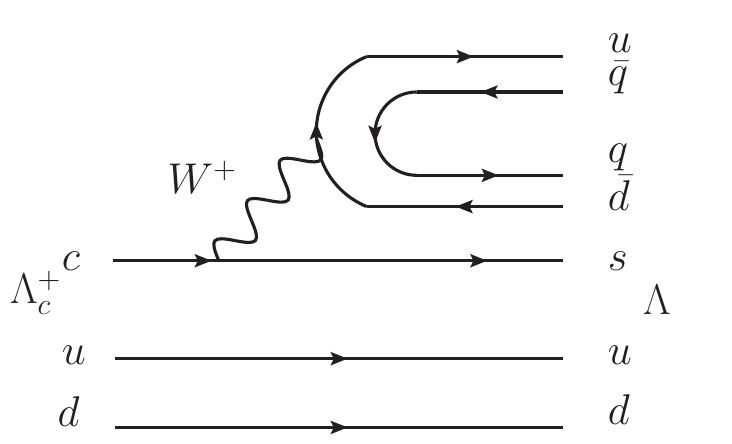}\label{fig:quark_level_b}
	}
	\subfigure[]{
		\centering
		\includegraphics[scale=0.33]{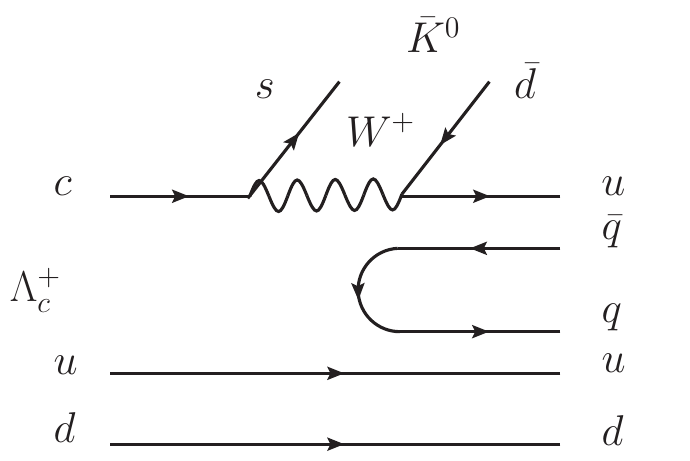}\label{fig:quark_level_c}
	}
	\subfigure[]{
		\centering
		\includegraphics[scale=0.33]{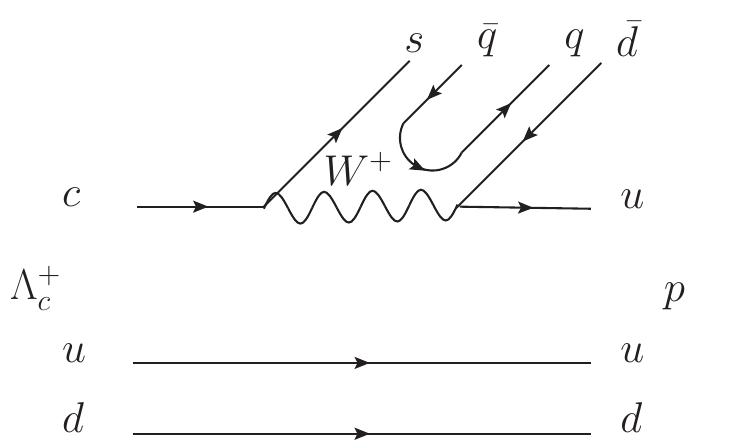}\label{fig:quark_level_d}
	}
	\caption{Quark level diagram for the process (a) $\Lambda_c^+\to\pi^+s(\bar{q}q)ud$ via $W^+$ external emission, (b)$\Lambda_c^+\to u(\bar{q}q)\bar{d}~\Lambda$ via $W^+$ external emission, (c) $\Lambda_c^+\to \bar{K}^0 u(\bar{q}q)\bar{d}$ via $W^+$ internal emission and (d) $\Lambda_c^+\to s(\bar{q}q)\bar{d}~p$ via $W^+$ internal emission.}\label{fig:quark_level}
\end{figure*}

\begin{figure}[htbp]
 \subfigure[]{
		\centering
		\includegraphics[scale=0.55]{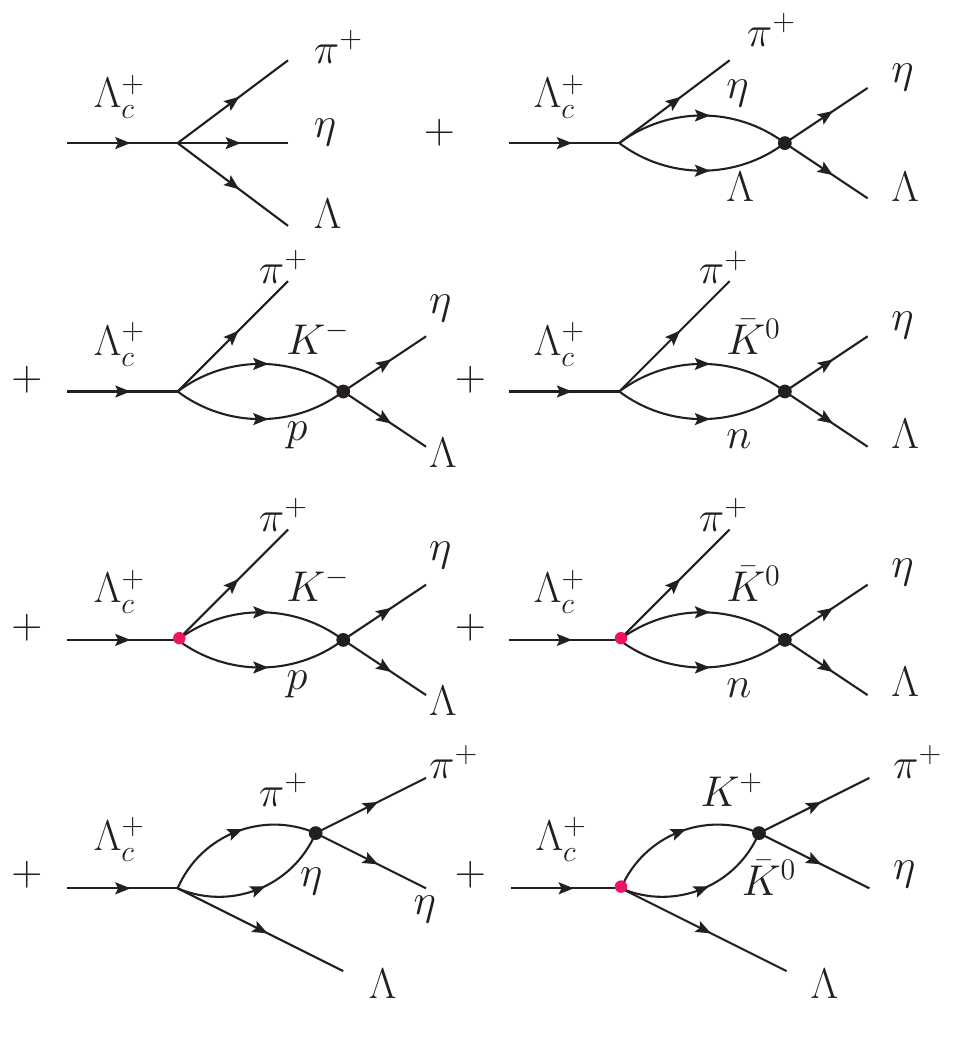}\label{fig:machenism_rescattering}
	}
	\subfigure[]{
		\centering
		\includegraphics[scale=0.65]{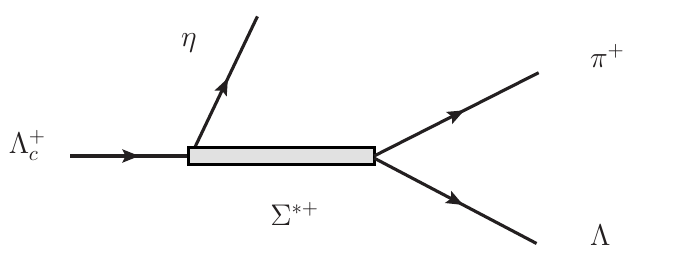}\label{fig:machenism_intermediate}
	}
	\caption{ Mechanisms for (a) tree level $\Lambda_c^+\to\eta\pi^+\Lambda$ and rescattering of intermediate components, (b) intermediate states.}\label{fig:machenism_process}
\end{figure}

It is interesting to investigate how the $\Lambda_c^+\to\eta\pi^+\Lambda$ reaction proceeds at the microscopic level. 
As shown in Fig.~\ref{fig:quark_level_a} for the $W^+$ external emission mechanism, by considering the hadronization of all quarks, alongside their flavor and spin wave functions~\cite{Duan:2024czu,Xie:2016evi}, we obtain
\begin{equation}
	\Lambda_c^+\Rightarrow\pi^+\left\{\frac{1}{\sqrt{2}}K^-p+\frac{1}{\sqrt{2}}\bar{K}^0n+\frac{1}{3}\eta\Lambda\right\}.
\end{equation}
Regarding Fig.~\ref{fig:quark_level_b}, a subtle point arises: as discussed in detail in Ref.~\cite{Duan:2024okk}, the $\eta\pi^+$ and $\pi^+\eta$ terms do not sum up but rather cancel each other out. 
This is due to the $[P,\partial_\mu P]W^\mu$ structure of the $WPP$ vertex~\cite{Gasser:1983yg,Scherer:2002tk,Ren:2015bsa}, where $P$ is the pseudoscalar meson matrix and $W^\mu$ represents the weak gauge boson field. 
Similarly, since $\mu=0$ provides the dominant contribution, the $K^+\partial_0\bar{K}^0-\bar{K}^0\partial_0 K^+$ term within the $K^+\bar{K}^0$ channel also yields no net contribution~\cite{Duan:2024okk}. 
Consequently, this diagram is omitted from the present study, which is consistent with Refs.~\cite{Duan:2024czu,Zhang:2026igc,Li:2025gvo}.

On the other hand, in Fig.~\ref{fig:quark_level_c} for the $W^+$ internal emission mechanism, considering the hadronization of all quarks along with their flavor and spin wave functions~\cite{Duan:2024czu,Li:2026lbo}, we have
\begin{equation}
	\Lambda_c^+\Rightarrow\frac{\bar{K}^0}{\sqrt{2}}\left\{\left(\frac{\pi^0}{\sqrt{2}}+\frac{\eta}{\sqrt{3}}\right)p+\pi^+n-\sqrt{\frac{2}{3}}K^+\Lambda\right\},
\end{equation}
where the final states $\bar{K}^0 \pi^+ n$ and $K^+\bar{K}^0\Lambda$ emerge. The $\bar{K}^0 n$ and $\bar{K}^0K^+$ pairs can subsequently interact to produce  $\eta\Lambda$ and $\pi^+\eta$, respectively.
Furthermore, when considering the hadronization in Fig.~\ref{fig:quark_level_d}, we have
\begin{equation}
	\Lambda_c^+\Rightarrow\frac{p}{\sqrt{2}}\left\{K^-\pi^+ -\frac{1}{\sqrt{2}}\bar{K}^0\pi^0\right\},
\end{equation}
where the final state $K^-p\pi^+$ emerges and the $K^-p$ pair can subsequently interact to produce $\eta\Lambda$.
It is noted that an additional parameter arises because the internal emission mechanism differs from the external one.

Based on the above discussions, we can rewrite the required hadronic pairs as follows, 
\begin{eqnarray}
\begin{aligned}\label{eq:hadronization}
	&\Lambda_c^+\Rightarrow\frac{1}{3}\pi^+\eta\Lambda+\frac{1}{\sqrt{2}}\pi^+K^-p+\frac{1}{\sqrt{2}}\pi^+\bar{K}^0n \\
   &+\gamma\left\{\frac{1}{\sqrt{2}}\pi^+\bar{K}^0n-\frac{1}{\sqrt{3}}\bar{K}^0K^+\Lambda+\frac{1}{\sqrt{2}}\pi^+K^-p\right\}.   
\end{aligned}
\end{eqnarray}
Here, $\gamma$ denotes the relative weight of internal versus external emissions at the quark level.
In principle, one may introduce different values of $\gamma$ for these two mechanisms, as shown in Figs.~\ref{fig:quark_level_c} and~\ref{fig:quark_level_d}. Nevertheless, because a common value of $\gamma$ already provides a satisfactory description of our results, we adopt the same relative weight for both mechanisms to reduce the number of free parameters.

The above production mechanisms at the quark level can be expressed at the hadron level as shown in Fig.~\ref{fig:machenism_rescattering}, where the tree-level process and subsequent rescattering processes are considered.
Furthermore, the $W^+$ internal emission vertices are marked with a red dot to indicate that there is a different coupling constant from the external emissions.\footnote{For the fifth and sixth internal emission mechanisms which stem from Figs.~\ref{fig:quark_level_c} and~\ref{fig:quark_level_d}, in the analysis of Ref.~\cite{Duan:2024czu} we shall also see that they have not been considered either, and they play a relevant role in this process.}. 
Generally, the amplitude in Fig.~\ref{fig:machenism_rescattering} can be expressed as
\begin{equation}\label{eq:t_tilde}
	\tilde{\mathcal{T}}=A \left(\mathcal{T}_{\text{Tree}} + \mathcal{T}_{MB} + \mathcal{T}_{MM}\right),
\end{equation}
with
\begin{equation}\label{eq:tree}
 \mathcal{T}_{\text{Tree}}= h_{\pi^+\eta\Lambda},
\end{equation}
\begin{eqnarray}\label{eq:MB}
 \mathcal{T}_{MB}&=&h_{\pi^+\eta\Lambda}G_{\eta\Lambda}(M_{\text{inv}}(\eta\Lambda))T_{\eta\Lambda,\eta\Lambda}(M_{\text{inv}}(\eta\Lambda)) \nonumber\\
	&&+h_{\pi^+\bar{K}N}G_{K^-p}(M_{\text{inv}}(\eta\Lambda))T_{K^-p,\eta\Lambda}(M_{\text{inv}}(\eta\Lambda))   \nonumber\\
	&&+h_{\pi^+\bar{K}N}G_{\bar{K}^0n}(M_{\text{inv}}(\eta\Lambda))T_{\bar{K}^0n,\eta\Lambda}(M_{\text{inv}}(\eta\Lambda)), \nonumber\\
    &&+\gamma h_{\pi^+\bar{K}N}G_{K^-p}(M_{\text{inv}}(\eta\Lambda))T_{K^-p,\eta\Lambda}(M_{\text{inv}}(\eta\Lambda))   \nonumber\\
	&&+\gamma h_{\pi^+\bar{K}N}G_{\bar{K}^0n}(M_{\text{inv}}(\eta\Lambda))T_{\bar{K}^0n,\eta\Lambda}(M_{\text{inv}}(\eta\Lambda)), \nonumber\\
\end{eqnarray}
\begin{eqnarray}\label{eq:MM}
 &\mathcal{T}_{MM}=h_{\pi^+\eta\Lambda}G_{\pi^+\eta}(M_{\text{inv}}(\pi^+\eta))T_{\pi^+\eta,\pi^+\eta}(M_{\text{inv}}(\pi^+\eta))  \nonumber\\
	&+\gamma h_{K^+\bar{K}^0\Lambda}G_{K^+\bar{K}^0}(M_{\text{inv}}(\pi^+\eta))T_{K^+\bar{K}^0,\pi^+\eta}(M_{\text{inv}}(\pi^+\eta)),\nonumber\\
\end{eqnarray}
where $A$ is a global normalization constant. 
According to the Eq.~\eqref{eq:hadronization}, the coefficients are:
\begin{equation}
	h_{\pi^+\eta\Lambda}=\frac{1}{3}, \quad h_{\pi^+\bar{K}N}=\frac{1}{\sqrt{2}}, \quad h_{K^+\bar{K}^0\Lambda}=-\frac{1}{\sqrt{3}}.
\end{equation} 
The transition amplitudes $T_{i,j}$ in the pseudoscalar meson-baryon scattering $t_{MB}$ in Eq.~(\ref{eq:MB}) are derived from the chiral unitary approach, incorporating the coupled channels $K^-p$, $\bar{K}^0n$, $\pi^+\Sigma^-$, $\pi^-\Sigma^+$, $\pi^0\Sigma^0$, $\pi^0\Lambda$, $\eta\Lambda$, $\eta\Sigma^0$, $K^+\Xi^-$, and $K^0\Xi^0$, which can dynamically generate the $\Lambda(1670)$ state as described in Refs.~\cite{Oset:1997it,Lyu:2024qgc,Duan:2024czu}. 
We employ dimensional regularization to evaluate the loop functions and set $\mu=630$~MeV~\cite{Oset:2001cn}. 
Since the pole position of the $\Lambda(1670)$ is highly sensitive to $a_{K\Xi}$ but moderately sensitive to others, we fix $a_{K\Xi}=-2.776$~\cite{Zhang:2024jby,Li:2026umb} and adopt $a_{\bar{K}N}=-1.84$, $a_{\pi\Lambda}=-1.83$, $a_{\eta \Sigma}=-2.38$, $a_{\pi\Sigma}=-2.00$, and $a_{\eta\Lambda}=-2.25$ from Ref.~\cite{Oset:2001cn}. 
The pole position calculated from this parameter set is $1669.0-20.3i$~MeV, which is consistent with the result of Ref.~\cite{Kamano:2015hxa}. 
While for the amplitudes in the pseudoscalar meson-pseudoscalar meson scattering $t_{MM}$, we consider two channels $K^+\bar{K}^0$ and $\pi^+\eta$ and apply the cutoff method to regularize the propagator with $q_{\text{max}}=600$~MeV~\cite{Lin:2021isc,Li:2025gvo,Lyu:2025oow}, in which the $a_0(980)$ state can be dynamically generated.
In this way, the line shapes of $a_0(980)$ and $\Lambda(1670)$ are fixed by the previous work without any free parameters.

Furthermore, the contributions from two intermediate states $\Sigma(1385)~(3/2^+)$ and $\Sigma(1380)~(1/2^-)$ are taken into account, which are illustrated in Fig.~\ref{fig:machenism_intermediate}. 
Specifically, the excitation of the $\Sigma(1385)$ involves a transition from a $1/2^+$ baryon state to a $3/2^+$ state.
Similar to the nucleon-$\Delta$ transition, this is described using the $S^\dagger$ transition spin operator, which is widely employed in the $\Delta$-hole model for pion nuclear physics~\cite{Ericson:1988gk,Oset:1981ih}. 
It is defined as~\cite{Ericson:1988gk,Oset:1981ih,Duan:2024czu,Lyu:2026rsm}:
\begin{equation}
	\left\langle\frac{3}{2}M\left|S^\dagger_\nu\right|\frac{1}{2}m\right\rangle=C\left(\frac{1}{2},\mathbf{1},\frac{3}{2}; m,\nu,M\right)\left\langle\frac{3}{2}\left|\left|S^\dagger\right|\right|\frac{1}{2}\right\rangle,
\end{equation}
where $\nu$ is the spherical component of the rank-$\mathbf{1}$ operator $S^\dagger$. 
This definition explicitly relies on the Wigner-Eckart theorem, with the reduced matrix element conventionally set to $\mathbf{1}$. 
For practical calculations, one uses:
\begin{equation}\label{eq:S}
	\sum_M S_l|M\rangle\langle M|S^\dagger_m=\frac{2}{3}\delta_{lm}-\frac{i}{3}\epsilon_{lms}\sigma_s,
\end{equation}
where $l$, $m$, and $s$ denote Cartesian components. 
The amplitude for the $\Sigma(1385)$ excitation, assuming a $P$-wave coupling at each $1/2\to3/2$ transition vertex, is given by:
\begin{eqnarray}
	\mathcal{T}_{\Sigma(1385)}&=&\frac{\beta}{M_{\Lambda}}\left\langle m^\prime\left|S_i\vec{P}^*_{\pi^+ i}\right|M\right\rangle\left\langle M\left|S_j^\dagger\vec{P}_{\eta j}^*\right|m\right\rangle \mathcal{D}  \nonumber \\
	&=&\frac{\beta}{M_{\Lambda}}\mathcal{D}\left\langle m^\prime\left|\frac{2}{3}\delta_{ij}-\frac{i}{3}\epsilon_{ijs}\sigma_s\right|m\right\rangle \vec{P}^*_{\pi^+ i}\vec{P}_{\eta j}^* \\
	&=&\frac{\beta}{M_{\Lambda}}\mathcal{D}\left\langle m^\prime\left|\frac{2}{3}\vec{P}^*_{\pi^+}\cdot\vec{P}_{\eta}^*-\frac{i}{3}\epsilon_{ijs}\sigma_s\vec{P}^*_{\pi^+ i}\vec{P}_{\eta j}^*\right|m\right\rangle,  \nonumber 
\end{eqnarray}
where $\beta$ is the relative weight of the $\Sigma(1385)$ state, and 
\begin{equation}
	\mathcal{D}=\dfrac{1}{M_{\text{inv}}(\pi^+\Lambda)-M_{\Sigma(1385)}+{i\Gamma_{\Sigma(1385)}}/{2}},
\end{equation}
with $M_{\Sigma(1385)}=1382.83~\text{MeV}$ and $\Gamma_{\Sigma(1385)}=36.2~\text{MeV}$.
Actually, such form is well consistent with the usual Rarita-Schwinger propagator~\cite{Rarita:1941mf} of the spin-$3/2$ particle.

As discussed in the introduction, in addition to the $\Sigma(1385)$, we also account for the $\Sigma(1380)$ state. 
Since the $\Sigma(1380)$ has not yet been experimentally confirmed, we adopt a simple BW form for its contribution:
\begin{equation}\label{eq:t1380}
	\mathcal{T}_{\Sigma(1380)} =\dfrac{V_{\Sigma(1380)}M_{\Sigma(1380)}}{M_{\text{inv}}(\pi^+\Lambda)-M_{\Sigma(1380)}+{i\Gamma_{\Sigma(1380)}}/{2}},
\end{equation}
where $V_{\Sigma(1380)}$ represents its relative coupling strength. 
To minimize the number of free parameters, we fix its mass at $M_{\Sigma(1380)}=1380$~MeV and its width at $\Gamma_{\Sigma(1380)}=120$~MeV as used in Refs.~\cite{Wu:2009nw,Wu:2009tu,Liu:2017hdx,Li:2024rqb,Lyu:2024qgc}.

Finally, the total amplitude for the $\Lambda_c^+\to\eta\pi^+\Lambda$ process is constructed as
\begin{equation}\label{eq:ttotal}
	\mathcal{T}_{\text{Total}} =A\left\{\mathcal{T}_{\text{Tree}}+\mathcal{T}_{MB}+\mathcal{T}_{MM}+\mathcal{T}_{\Sigma(1385)}+\mathcal{T}_{\Sigma(1380)}\right\}.
\end{equation}
In order to study the influence of $\Sigma(1380)$, we consider two distinct scenarios in our calculations: Case I excludes the $\Sigma(1380)$ contribution, while Case II includes it. 
Since the free parameters $\beta$, $\gamma$ and $V_{\Sigma(1380)}$ involved in these cases are all complex numbers, the parameterizations are defined as follows
\begin{equation}\label{eq:twocase}
\begin{aligned}
&\text{I}:~&\beta \equiv \beta e^{i\phi}, 
~&\gamma \equiv \gamma e^{i\phi'};~& \\
&\text{II}:~&\beta \equiv \beta e^{i\phi},~&\gamma \equiv \gamma e^{i\phi'},~&V_{\Sigma(1380)} \equiv V_{\Sigma(1380)} e^{i\phi''}. 
\end{aligned}
\end{equation}

The analytical expression for the double differential width can be obtained by
\begin{equation}
	\dfrac{d^2\Gamma}{dM_{12}dM_{23}} = \frac{M_{12}~M_{23}}{(2\pi)^3}\dfrac{M_\Lambda }{2M_{\Lambda_c^+}^2}\overline{\sum}|\mathcal{T}_{\text{Total}}|^2.
\end{equation}
The subscripts $1$, $2$, and $3$ represent particles $\pi^+$, $\eta$, and $\Lambda$, respectively. 
In addition, the double differential width in terms of angular distribution and invariant mass can be rewritten as
\begin{equation}\label{Angle}
	\dfrac{d^2\Gamma}{dM_{\pi^+\Lambda}d{\rm cos}\theta} = \frac{|\vec{p}_{\eta}||\vec{p}^{\,*}_{\pi^+}|}{(2\pi)^3}\dfrac{M_{\Lambda}}{2M_{\Lambda_c^+}}\overline{\sum}|\mathcal{T}_{\text{Total}}|^2,
\end{equation}
where $\theta$ is the angle of the particles $\pi^+$ and $\eta$ in the center-of-mass frame of the $\pi^+\Lambda$ system. 
$\vec{p}_\eta$ is the three-momentum of $\eta$ meson in the rest frame of $\Lambda_c^+$, and $\vec{p}^{\,*}_{\pi^+}$ is the three-momentum of $\pi^+$ in the rest frame of the $\pi^+\Lambda$.

\section{Results and Discussions}\label{sec3}

\begin{table*}
\centering
\caption{Free parameter values fitted to the MC sample with 829 data points.}
\begin{tabular}{|c|c|c|c|c|c|c|c|c|}
	\hline
	~ & $A$ & $\beta$ & $\gamma$ & $\phi$ & $\phi^\prime$ &$V_{\Sigma(1380)}$ & $\phi^{\prime\prime}$ & $\chi^2 / \text{d.o.f.}$ \\
	\hline
	w/o $\Sigma(1380)$ &~~$1.52$~MeV$^{-1}$~~&~~$0.23$~~&~~$0.76$~~&~~$0.25\pi$~~&~~$1.42\pi$~~&~~-~~&~~-~~&~~$8.75$~~\\
    w $\Sigma(1380)$ &~~$1.41$~MeV$^{-1}$~~&~~$0.25$~~&~~$0.94$~~&~~$0.37\pi$~~&~~$1.40\pi$~~&~~$0.01$~~&~~$-0.45\pi$~~&~~$2.92$~~\\
	\hline
\end{tabular}
\label{tab:MC_parameters}
\end{table*}

\begin{figure*}
	\subfigure[]{
		\centering
		\includegraphics[scale=0.65]{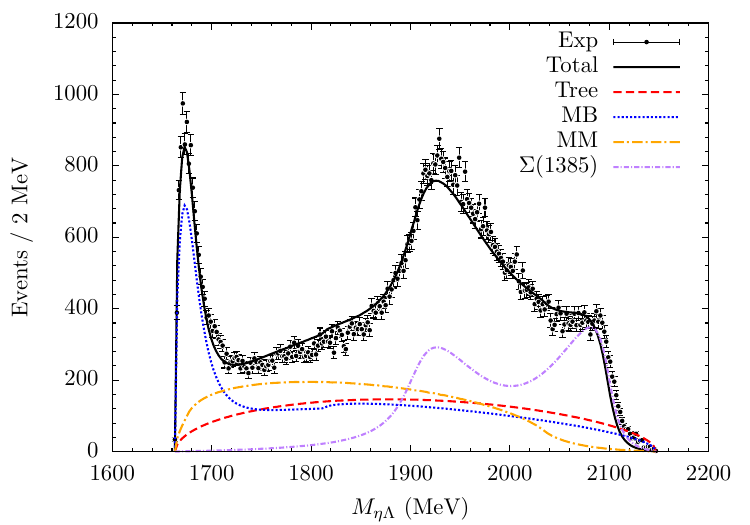}
	}
	\subfigure[]{
		\centering
		\includegraphics[scale=0.65]{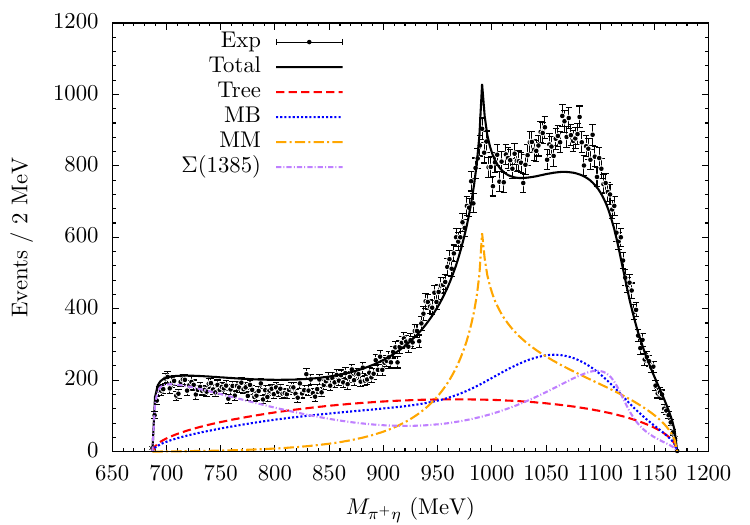}
	}
	\subfigure[]{
		\centering
		\includegraphics[scale=0.65]{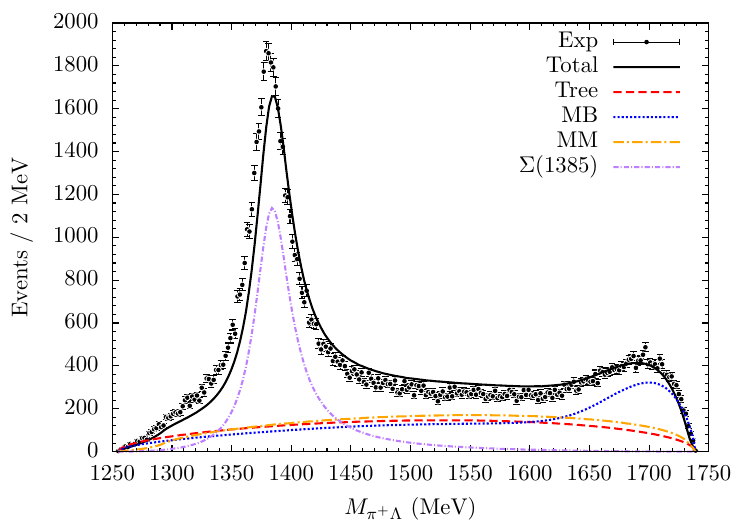}
	}
    \subfigure[]{
		\centering
		\includegraphics[scale=0.65]{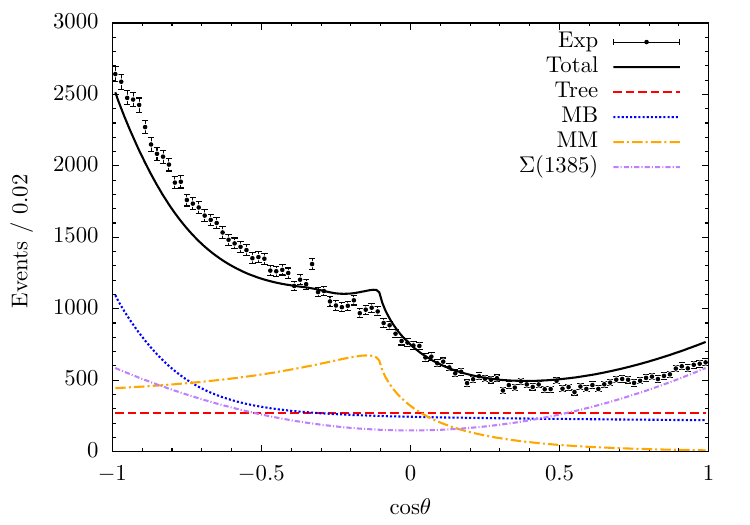}
	}
	\caption{$\eta\Lambda$ (a), $\pi^+\eta$ (b), $\pi^+\Lambda$ (c) invariant mass distributions and angular distribution (d) for the $\Lambda_c^+\to\eta\pi^+\Lambda$ decay without the contribution of the $\Sigma(1380)$. The BESIII MC data are provided by the BESIII Collaboration.}\label{fig:MC_distribution_without1380}
\end{figure*}

\begin{figure*}
 \subfigure[]{
		\centering
		\includegraphics[scale=0.65]{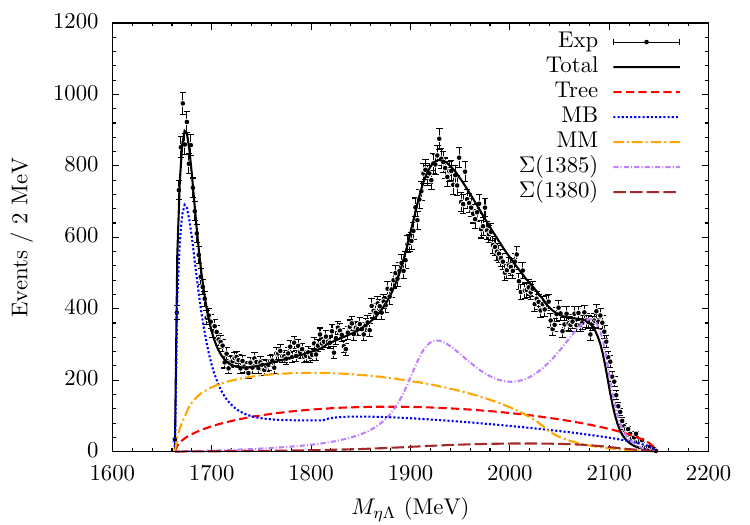}
	}
	\subfigure[]{
		\centering
		\includegraphics[scale=0.65]{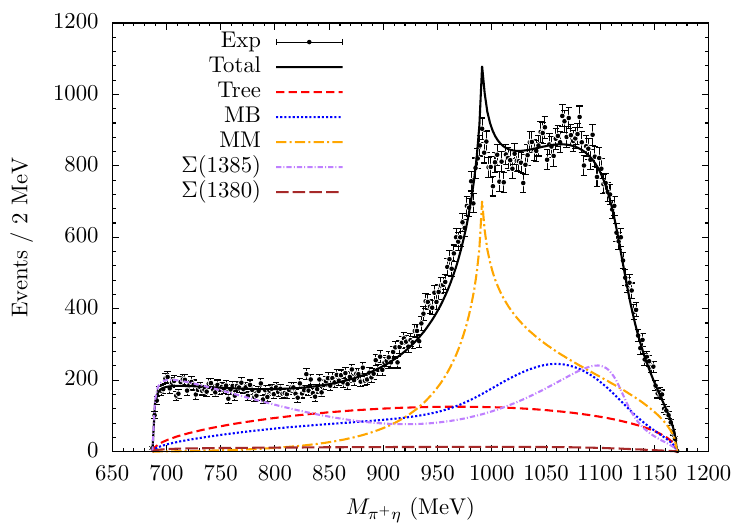}
	}
	\subfigure[]{
		\centering
		\includegraphics[scale=0.65]{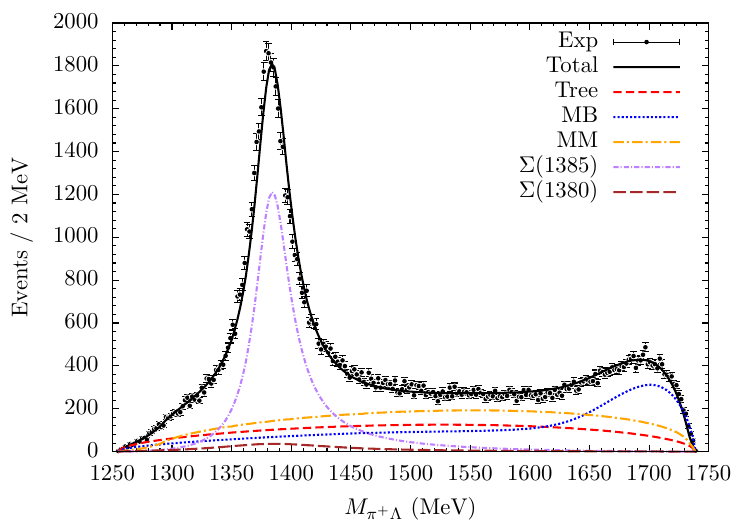}
	}
    \subfigure[]{
		\centering
		\includegraphics[scale=0.65]{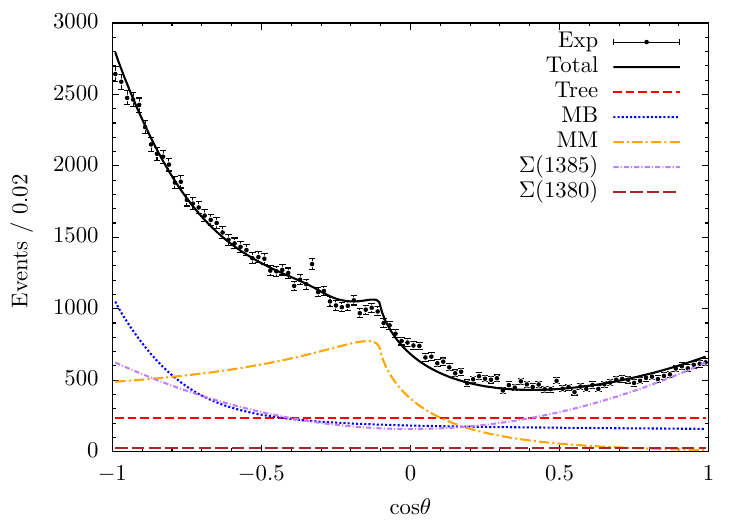}
	}
	\caption{$\eta\Lambda$ (a), $\pi^+\eta$ (b), $\pi^+\Lambda$ (c) invariant mass distributions and angular distribution (d) for the $\Lambda_c^+\to\eta\pi^+\Lambda$ decay with the contribution of the $\Sigma(1380)$. The BESIII MC data are provided by the BESIII Collaboration.}\label{fig:MC_distribution_with1380}
\end{figure*}
\begin{figure*}
 \subfigure[]{
		\centering
		\includegraphics[scale=0.65]{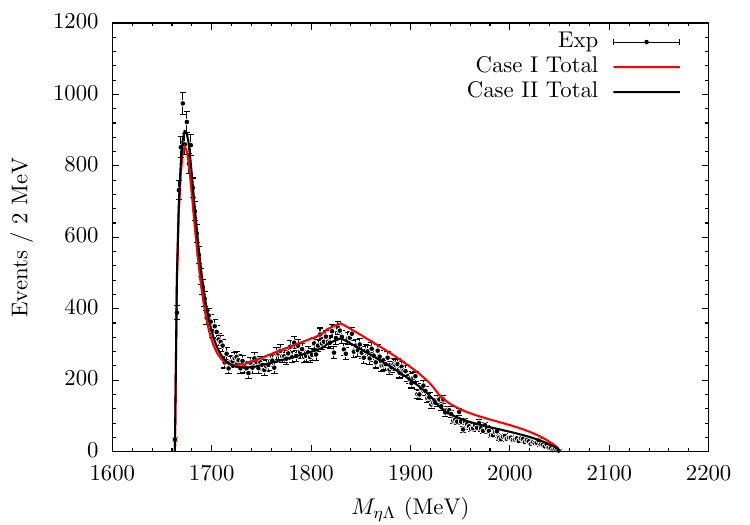}
	}
	\subfigure[]{
		\centering
		\includegraphics[scale=0.65]{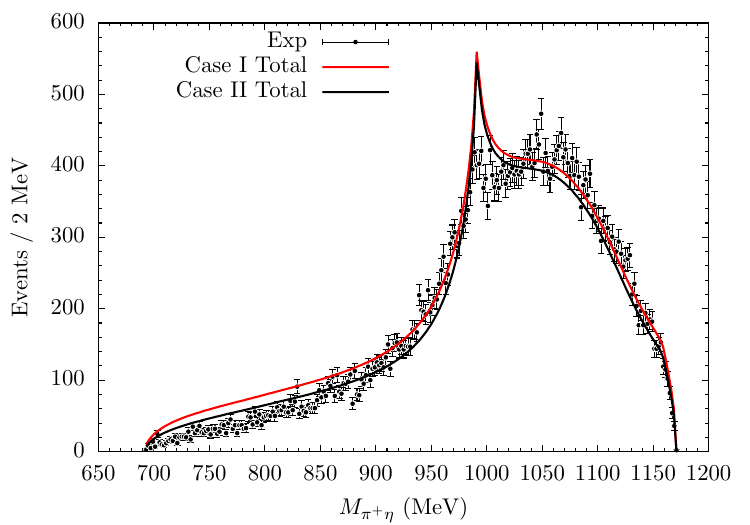}
	}
	\subfigure[]{
		\centering
		\includegraphics[scale=0.40]{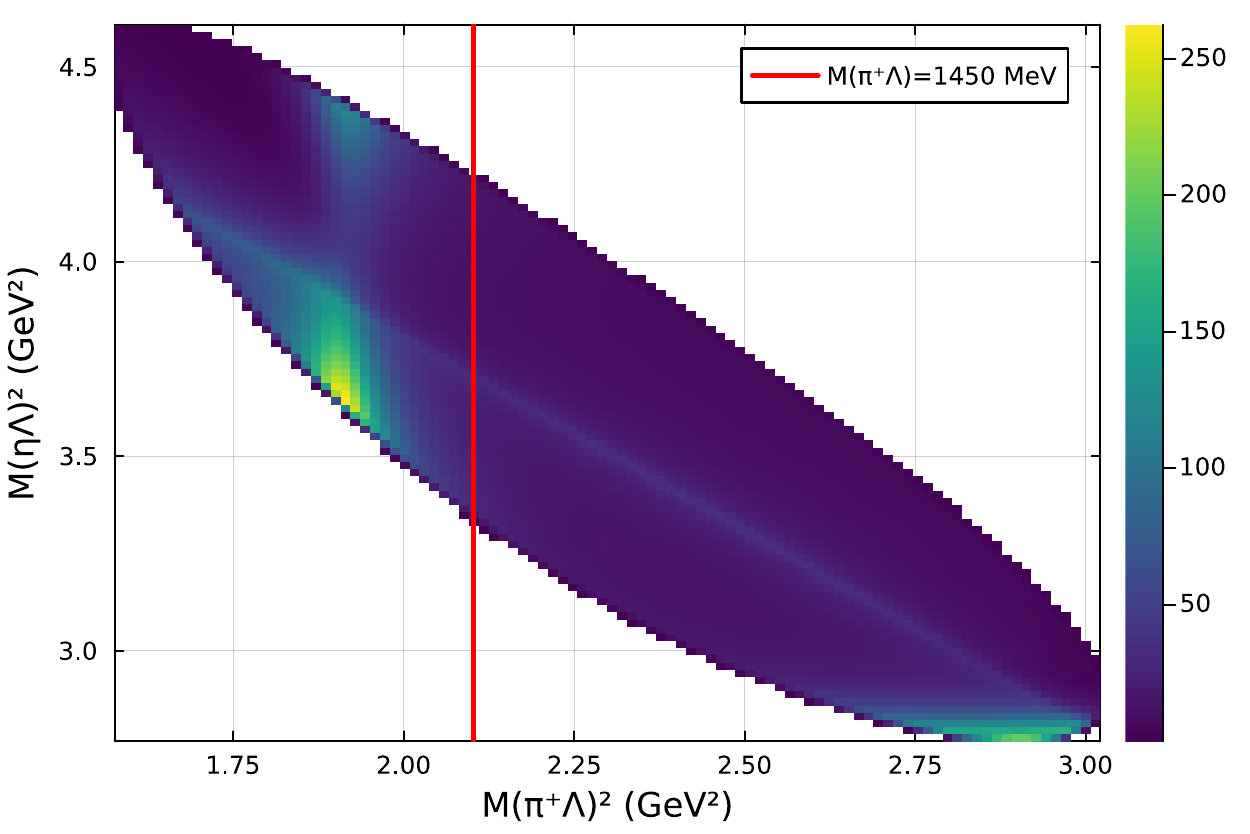}
	}
	\subfigure[]{
		\centering
		\includegraphics[scale=0.65]{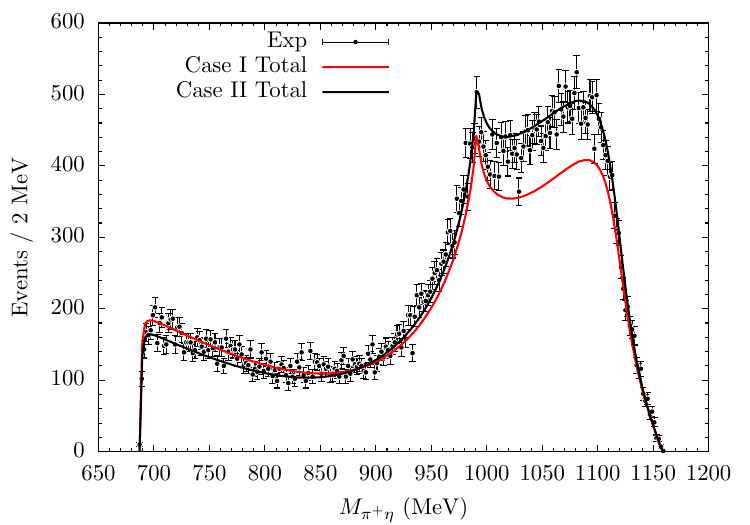}
	}
	\caption{Invariant mass distributions of $\eta\Lambda$ (a) and $\pi^+\eta$ (b) for the $\Lambda_c^+\to\eta\pi^+\Lambda$ decay with $M_{\pi^+\Lambda}>1450~\mathrm{MeV}$. The $\pi^+\Lambda$ versus $\eta\Lambda$ Dalitz plot with $\Sigma(1380)$ (c). The $\pi^+\eta$ invariant mass distribution (d) is shown for the region $M_{\pi^+\Lambda}<1450~\mathrm{MeV}$. The BESIII MC data are provided by the BESIII Collaboration.}\label{fig:MC_distribution_cut}
\end{figure*}
\begin{table*}
\centering
\caption{Free parameter values fitted to the Belle data sample.}
\begin{tabular}{|c|c|c|c|c|c|c|c|c|}
	\hline
	~ & $A$ & $\beta$ & $\gamma$ & $\phi$ & $\phi^\prime$ &$V_{\Sigma(1380)}$ & $\phi^{\prime\prime}$ & $\chi^2 / \text{d.o.f.}$ \\
	\hline
	w/o $\Sigma(1380)$ &~~$1.35$~~&~~$0.24$~~&~~$1.57$~~&~~$-0.05\pi$~~&~~$-1.05\pi$~~&~~-~~&~~-~~&~~$4.88$~~\\
    w $\Sigma(1380)$ &~~$1.18$~~&~~$0.28$~~&~~$1.77$~~&~~$0.05\pi$~~&~~$-0.98\pi$~~&~~$0.01$~~&~~$-0.62\pi$~~&~~$3.90$~~\\
	\hline
\end{tabular}
\label{tab:Belle_parameters}
\end{table*}

\begin{figure*}
	\subfigure[]{
		\centering
		\includegraphics[scale=0.65]{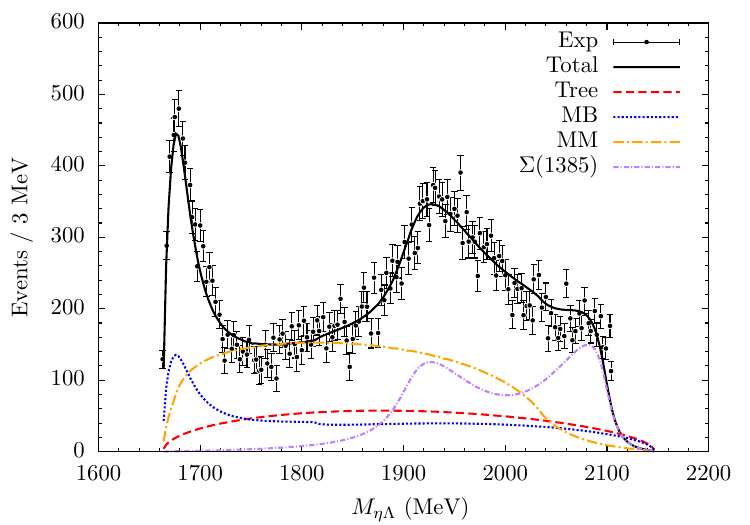}
	}
	\subfigure[]{
		\centering
		\includegraphics[scale=0.65]{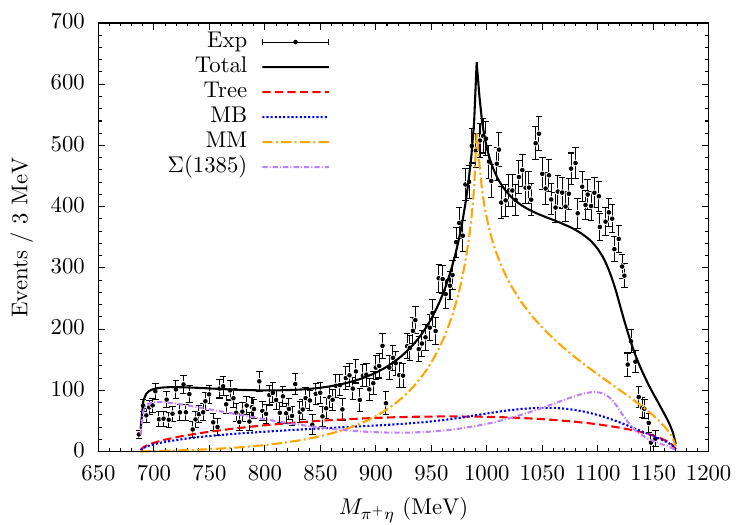}
	}
	\subfigure[]{
		\centering
		\includegraphics[scale=0.65]{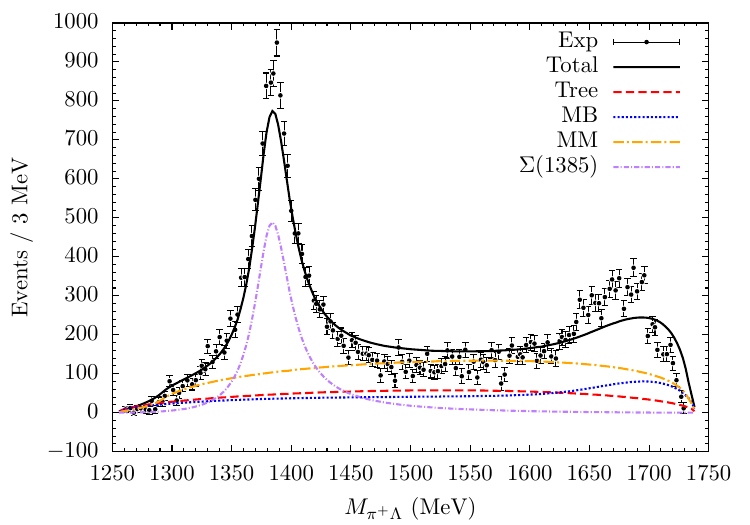}
	}
    \subfigure[]{
		\centering
		\includegraphics[scale=0.65]{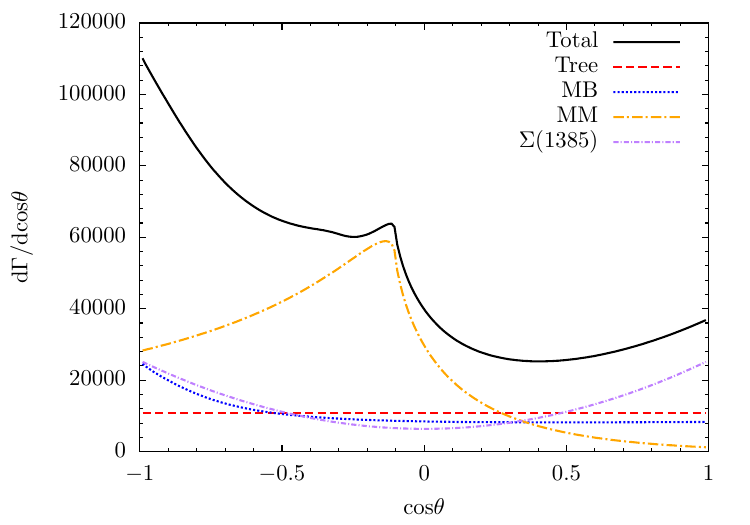}
	}
	\caption{$\eta\Lambda$ (a), $\pi^+\eta$ (b), $\pi^+\Lambda$ (c) invariant mass distributions and angular distribution (d) for the $\Lambda_c^+\to\eta\pi^+\Lambda$ decay without the contribution of the $\Sigma(1380)$. The experimental data are taken from the Belle Collaboration~\cite{Belle:thesis}, from which the background contribution has been subtracted.}\label{fig:Belle_distribution_without1380}
\end{figure*}

\begin{figure*}
 \subfigure[]{
		\centering
		\includegraphics[scale=0.65]{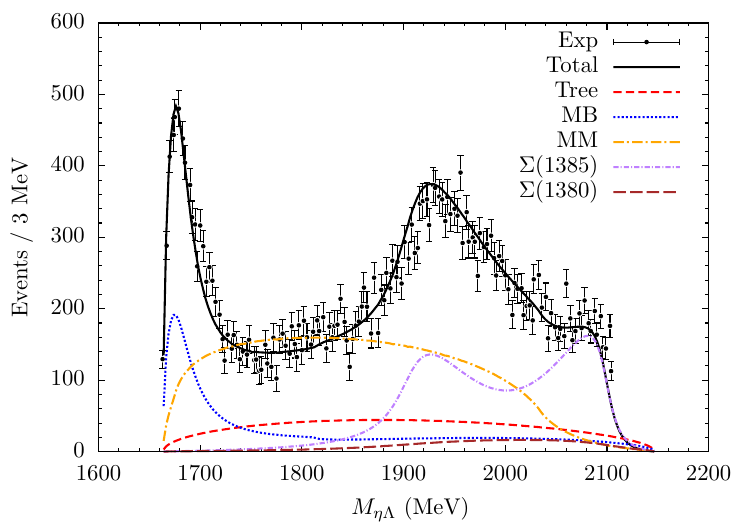}
	}
	\subfigure[]{
		\centering
		\includegraphics[scale=0.65]{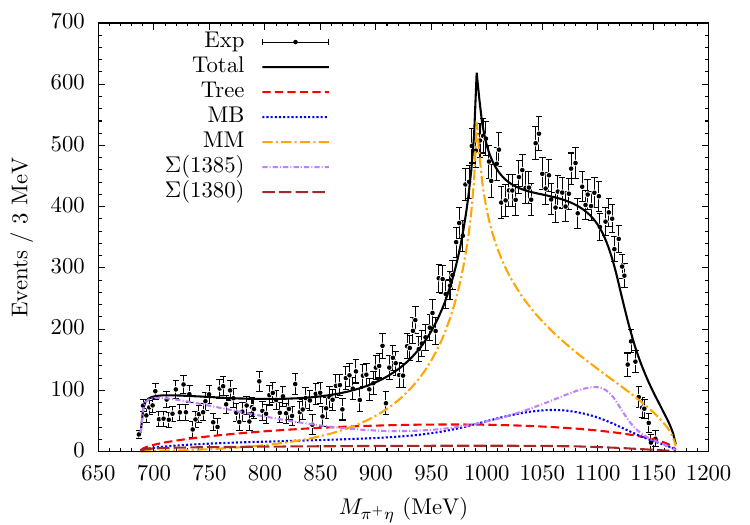}
	}
	\subfigure[]{
		\centering
		\includegraphics[scale=0.65]{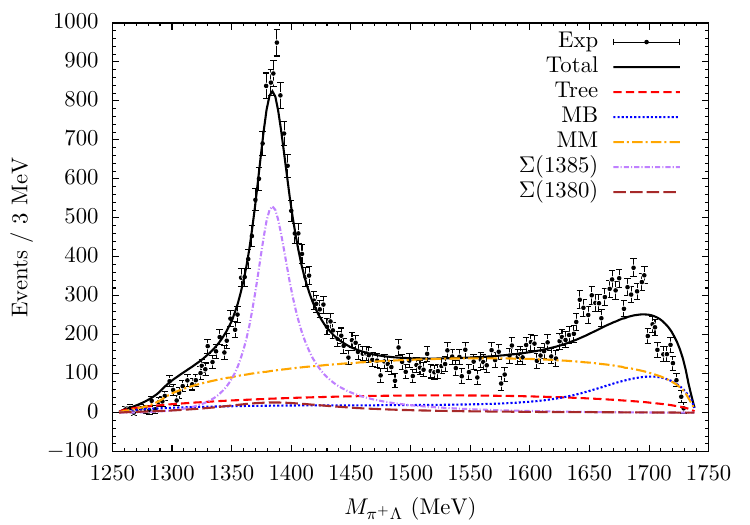}
	}
    \subfigure[]{
		\centering
		\includegraphics[scale=0.65]{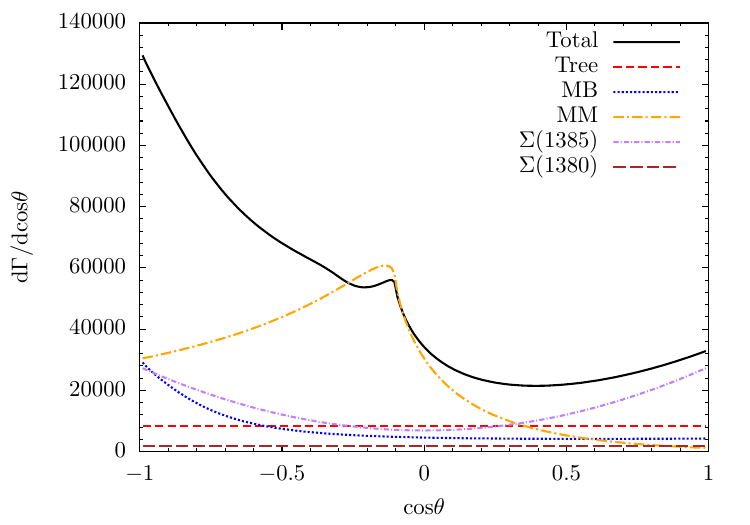}
	}
	\caption{$\eta\Lambda$ (a), $\pi^+\eta$ (b), $\pi^+\Lambda$ (c) invariant mass distributions and angular distribution (d) for the $\Lambda_c^+\to\eta\pi^+\Lambda$ decay with the contribution of the $\Sigma(1380)$. The experimental data are taken from the Belle Collaboration~\cite{Belle:thesis}, from which the background contribution has been subtracted.}\label{fig:Belle_distribution_with1380}
\end{figure*}

In this work, we consider two scenarios. 
Case I does not include the $\Sigma(1380)$ contribution and contains five free parameters, namely $A$, $\beta$, $\gamma$, $\phi$, and $\phi^\prime$. 
Case II includes the $\Sigma(1380)$ contribution and involves seven free parameters: $A$, $\beta$, $\gamma$, $\phi$, $\phi^\prime$, $V_{\Sigma(1380)}$, and $\phi^{\prime\prime}$. 

To constrain these parameters, we first employ the MC sample provided by the BESIII Collaboration for the decay $\Lambda_c^+ \to \eta \pi^+ \Lambda$. 
Compared with the distributions reported in the published BESIII~\cite{BESIII:2024mbf} and Belle~\cite{Belle:2020xku} analyses, this MC sample contains more detailed kinematic information with background subtraction and efficiency corrections, and therefore allows for a more detailed test for the theoretical model. 
Specifically, the MC sample contains $10^5$ events.
From this sample, we extract the invariant mass distributions of the $\eta\Lambda$, $\pi^+\eta$, and $\pi^+\Lambda$ systems, together with the angular distribution in $\cos \theta$. 
The bin widths are set to $2\text{ MeV}$ for the invariant mass distributions and $0.02$ for the angular distribution. 
These observables provide complementary constraints on the model parameters and enable a more stringent comparison between the theoretical distributions and the MC simulation.

We then perform two fits to the BESIII MC sample, corresponding to Case I (w/o $\Sigma(1380)$) and Case II (w $\Sigma(1380)$). 
The fitted parameters are summarized in Table~\ref{tab:MC_parameters}, from which we can find the inclusion of the $\Sigma(1380)$ contribution leads to a significant improvement in the value of $\chi^2/\text{d.o.f.}$. 
Notably, the fitted magnitude of $\gamma$ is less than $1$, consistent with the expectation that the internal emission mechanism is color-suppressed relative to the external emission mechanism.

Fig.~\ref{fig:MC_distribution_without1380} shows the fitted $\eta\Lambda$, $\pi^+\eta$, and $\pi^+\Lambda$ invariant mass distributions, together with the angular distribution, obtained without including the $\Sigma(1380)$ contribution. 
The corresponding results including the $\Sigma(1380)$ contribution are shown in Fig.~\ref{fig:MC_distribution_with1380}. 
Here, the purple dash-dotted curves represent the contribution from the $\Sigma(1385)$, which produces a clear peak in the $\pi^+\Lambda$ invariant mass distribution. 
The orange dash-dotted curves show the contribution from the meson-meson interaction, which dynamically generates the scalar meson $a_0(980)$ associated with the cusp around $980\text{ MeV}$ in the $\pi^+\eta$ invariant mass distribution. 
The blue dotted curves indicate the contribution from the meson-baryon interaction, which dynamically generates the $\Lambda(1670)$ state associated with the threshold enhancement in the $\eta\Lambda$ invariant mass distribution. 
In Fig.~\ref{fig:MC_distribution_with1380}, the brown dashed curves represent the $\Sigma(1380)$ contribution. 
Finally, the red dashed curves represent the tree-level contribution, while the black solid curves show the total distribution obtained from the coherent sum of all amplitudes. 
It is evident that the inclusion of the $\Sigma(1380)$ contribution significantly improves the description of the distributions, particularly in the $1000\sim 1100 \text{ MeV}$ region of the $\pi^+\eta$ mass distribution, the $1300\sim 1400\text{ MeV}$ region of the $\pi^+\Lambda$ distribution, and the angular distribution. 
This improvement is obtained even though the fitted strength of the $\Sigma(1380)$ contribution remains relatively small. 
Furthermore, these results highlight that meson-meson and meson-baryon final-state interactions are crucial for accurately describing the line shapes of the $a_0(980)$ and $\Lambda(1670)$ states, respectively.

For Fig.~\ref{fig:MC_distribution_cut}(a) and Fig.~\ref{fig:MC_distribution_cut}(b), we select events satisfying $m_{\pi^+\Lambda}>1450~\mathrm{MeV}$ to suppress the contributions from $\Sigma(1385)$ and $\Sigma(1380)$. As indicated by the Dalitz plot of Fig.~\ref{fig:MC_distribution_cut}(c), the remaining structures in this selected region are mainly associated with the $\Lambda(1670)$ and $a_0(980)$ contributions. This region therefore provides useful information for constraining the relative weights of these states. Accordingly, we show the corresponding $\eta\Lambda$ and $\pi^+\eta$ invariant mass distributions, where both Case I and Case II are found to describe the data well. For Fig.~\ref{fig:MC_distribution_cut}(d), we further present the $\pi^+\eta$ invariant mass distribution by retaining events with $m_{\pi^+\Lambda}<1450~\mathrm{MeV}$, i.e., by excluding the complementary region $m_{\pi^+\Lambda}>1450~\mathrm{MeV}$. In this low-$m_{\pi^+\Lambda}$ region, a clear difference between Case I and Case II is observed around $1000\sim1100~\mathrm{MeV}$, indicating that the $\Sigma(1380)$ contribution plays an important role in describing the BESIII Monte Carlo data.

Next, we perform the same analysis using the background-subtracted Belle data~\cite{Belle:thesis}. 
As in the BESIII-MC analysis, we compare the fits with and without the $\Sigma(1380)$ contribution, and the corresponding fitted parameters are listed in Table~\ref{tab:Belle_parameters}. 
Apart from $\gamma$, the fitted parameters remain consistent with those obtained from the BESIII MC sample, and the inclusion of the $\Sigma(1380)$ again reduces $\chi^2/\text{d.o.f.}$. 

In contrast to the BESIII-MC fit, the Belle fit gives $\gamma > 1$ both with and without the $\Sigma(1380)$ contribution. 
Although this appears to deviate from the naive color suppression expectation, it can be attributed to the treatment of the Belle distributions in the present fit. 
In the Belle fit, these distributions are fitted directly without implementing the detector-efficiency and acceptance corrections. 
The missing experimental effects may modify the observed phase-space distributions and can be partly absorbed into the effective fit parameters. 
Therefore, a more quantitative determination of the internal emission strength would require a dedicated analysis incorporating the full detector efficiency and acceptance, such as with high-statistics Belle~II data.

Figs.~\ref{fig:Belle_distribution_without1380} and \ref{fig:Belle_distribution_with1380} illustrate the corresponding Belle fits without and with the $\Sigma(1380)$ contribution, respectively. 
The line styles and color assignments are the same as those used in the BESIII-MC analysis. 
The most visible changes induced by the $\Sigma(1380)$ contribution occur in the $1000\sim1130\text{ MeV}$ region of the $\pi^+\eta$ invariant mass distribution, the $1300\sim1350\text{ MeV}$ region of the $\pi^+\Lambda$ invariant mass distribution, and the whole region of the angular distribution. 
Therefore, these regions are important for determining the $\Sigma(1380)$ signal.

\section{Conclusions}\label{sec4}
In this work, we have performed a comprehensive theoretical study of the $\Lambda_c^+ \to \eta \Lambda\pi^+ $ decay to investigate the possible role of the $\Sigma(1380)$ state with $J^P=1/2^-$. 
Our theoretical framework incorporates the tree-level contribution, meson-baryon and meson-meson final-state interactions in coupled channels, and the intermediate $\Sigma(1385)$ and $\Sigma(1380)$ resonances. 
Within this approach, the $\Lambda(1670)$ and $a_0(980)$ states are dynamically generated from the meson-baryon and meson-meson interactions, respectively, providing a natural explanation for the corresponding structures in the invariant mass distributions.

By comparing our model predictions with the MC sample from the BESIII Collaboration and the experimental data from the Belle Collaboration, we find that the inclusion of the $\Sigma(1380)$ contribution substantially improves the description of the relevant distributions. 
Specifically, incorporating the $\Sigma(1380)$ contribution reduces the value of $\chi^2/\text{d.o.f.}$ in both fits and allows the model to better reproduce the experimental data. 
Furthermore, our analysis identifies several specific kinematic regions that are particularly sensitive to the possible $\Sigma(1380)$ contribution. 
These include the $1000 \sim 1130\text{ MeV}$ region of the $\pi^+\eta$ invariant mass distribution, the $1300 \sim 1350\text{ MeV}$ region of the $\pi^+\Lambda$ invariant mass distribution, and the whole region of the angular distribution.

Therefore, with Belle~II now accumulating a substantially larger data sample, a combined Belle and Belle~II analysis of this decay would be highly valuable. 
Such an effort could provide a more stringent test of the possible $\Sigma(1380)$ state and further clarify the spectrum of low-lying excited hyperons in the isovector $I=1$ sector.
Furthermore, the present theoretical model provides a useful parametrization for future experimental data analyses, as it involves only 7 (or 5) free parameters, compared with the 18 free parameters used in Ref.~\cite{BESIII:2024mbf}. Meanwhile, we suggest that this model can be employed in future experimental analyses, since it provides a reliable description of hadronic resonances with strong coupled-channel dynamics, including the $a_0(980)$ and $\Lambda(1670)$. This may help reduce systematic uncertainties in extracting resonance information from the data.

\section*{Acknowledgments}
We acknowledge Xiao-Rui Lyu and Xu-Dong Yu for fruitful discussions. We acknowledge the BESIII Collaboration for providing the Monte Carlo data used in this work.
This work was supported by the National Key R\&D Program of China (Grant Nos. 2024YFE0105200, 2025YFA1613900), the Natural Science Foundation of Henan (Grant No. 232300421140 and No. 252300423951), the National Natural Science Foundation of China (Grant Nos. 12475086, 12221005, and 12192263), the Chinese Academy of Sciences under Grant No. YSBR-101, and the Zhengzhou University Young Student Basic Research Projects for PhD students (Grant No. ZDBJ202522). Wen-Tao Lyu acknowledges the support of the China Scholarship Council.

\end{document}